\documentclass[preprint,12pt]{elsarticle}

\usepackage{amssymb}
\usepackage{amsmath}
\usepackage{hyperref}
\newcommand\ra{\rangle}
\newcommand\la{\langle}

\newcommand{\non}{\nonumber\\}

\newcommand{\be}{\begin{equation}}
\newcommand{\ee}{\end{equation}}
\newcommand{\bea}{\begin{eqnarray}}
\newcommand{\eea}{\end{eqnarray}}

\usepackage{color}

\usepackage[normalem]{ulem}  % \sout{old text} for strikeout

\begin{document}

\begin{frontmatter}



\title{Nuclear chiral density wave in neutron stars?}

%Title suggestion: Is there a chiral density wave in the interior of neutron stars?


\author[a]{Orestis Papadopoulos} %% Author name
\author[a]{Andreas Schmitt} %% Author name

\affiliation[a]{%Department and Organization
            addressline={Mathematical Sciences and STAG Research Centre, University of Southampton, Highfield Campus},
            city={Southampton},
            postcode={SO17 1BJ},
            country={United Kingdom}}

%% Abstract
\begin{abstract}
Anisotropic phases potentially play a role in the internal composition of neutron stars, the main laboratory for the phase structure of QCD at high baryon densities. We review the study of such a phase, the chiral density wave, within a phenomenological nucleon-meson model, including nucleonic vacuum fluctuations within a renormalization scheme recently developed. Neutron stars in this model and within our approximations either do not contain a chiral density wave core or they are too light to agree with observations.
\end{abstract}

%% Keywords
\begin{keyword}
neutron stars \sep nuclear matter \sep chiral density wave  \sep QCD 


\end{keyword}

\end{frontmatter}



\section{Introduction: Chiral density wave in dense matter}
\label{sec1}

\begin{figure}
\begin{center}
\includegraphics[width=0.7\textwidth]{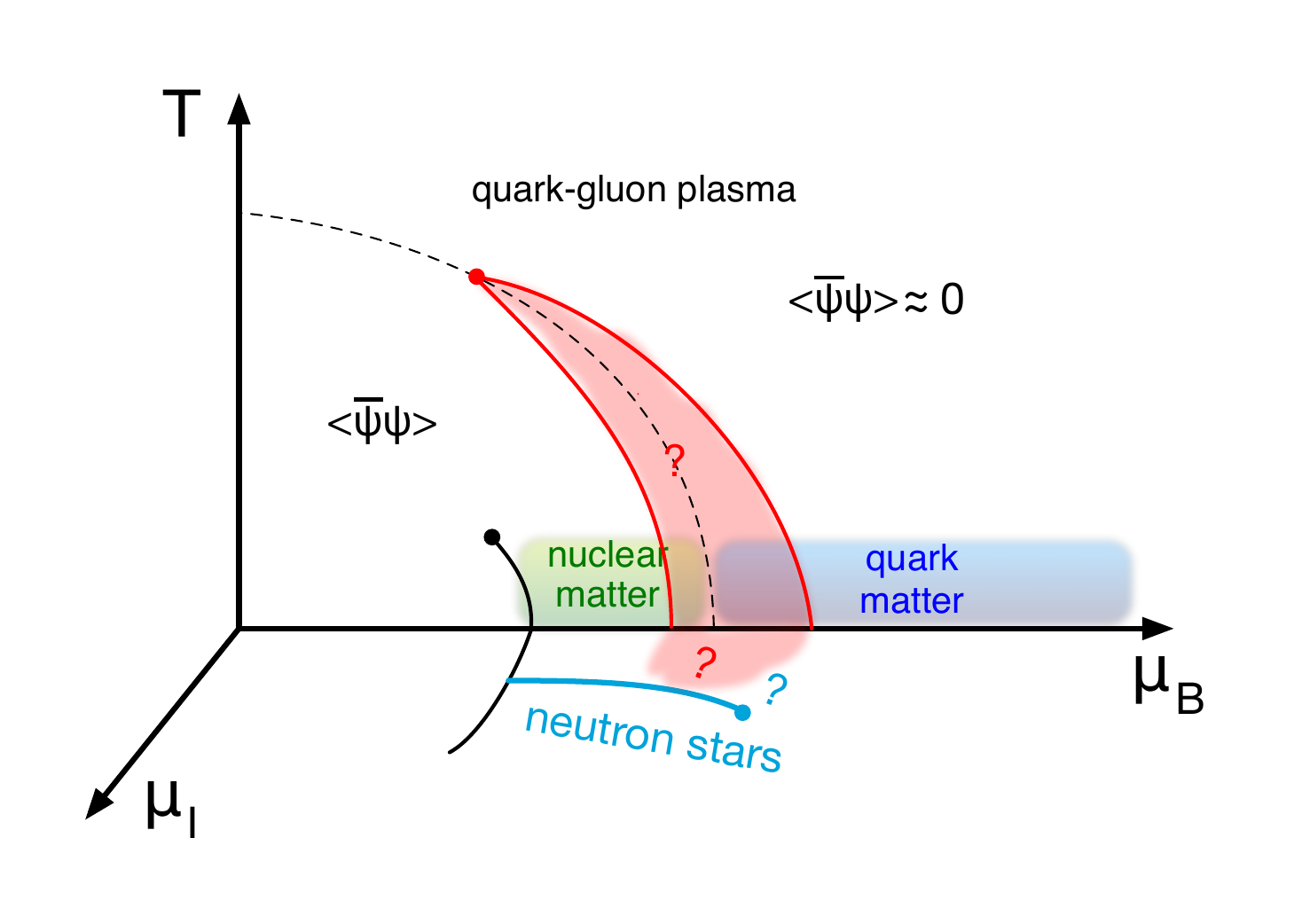}
\caption{Schematic view of the QCD phase diagram in the space spanned by temperature $T$, baryon chemical potential $\mu_B$, and isospin chemical potential $\mu_I$. It is conceivable that an anisotropic or inhomogeneous phase such as the CDW, is favored in the vicinity of the chiral phase transition (red area). In these proceedings we identify the CDW region at $T=0$ in the parameter space of a nucleon-meson model and test whether realistic neutron stars may have a CDW core.}   
\label{fig:QCDpd}
\end{center}
\end{figure}

The study of neutron star properties is closely tied to the physics of  dense matter governed by Quantum Chromodynamics (QCD). It is an open question whether the densities in neutron stars are large enough for  quark matter to exist in its core, although some indications for a deconfinement transition have been pointed out \cite{Annala:2019puf}. A transition between nuclear and quark matter inside the star would provide a natural environment for a spatially inhomogeneous mixed phase \cite{Schmitt:2020tac}, not unlike the nuclear pasta phases conjectured in the crust-core transition of the star \cite{Ravenhall:1983uh}. Besides this conceptually straightforward phase separation, more subtle effects that break rotational and/or translational invariance on a microscopic level are possible  \cite{Alford:2000ze,Buballa:2014tba}. Thermal fluctuations tend to  disfavor such phases, but isolated neutron stars have essentially zero temperature compared to the relevant baryon chemical potentials. Similarly, one may ask whether neutron star conditions such as electric charge neutrality and equilibrium with respect to the weak interactions tend to favor or disfavor anisotropic phases, see Fig.\ \ref{fig:QCDpd}. We shall address this question for the chiral density wave (CDW), a certain anisotropic phase that is particularly simple to implement theoretically and that is conjectured to exist in the vicinity of the chiral phase transition \cite{Nickel:2009wj}.

The CDW is induced by a chiral condensate that rotates periodically between the scalar and pseudoscalar sectors along a certain spatial direction. Unless rotational symmetry is already broken explicitly, for example by a magnetic field, the CDW breaks rotational symmetry spontaneously. The CDW has been studied before within phenomenological models, mainly based on quark degrees of freedom, like the Nambu--Jona-Lasinio model \cite{Nakano:2004cd} or the quark-meson model \cite{Carignano:2014jla}, and first applications to neutron stars were considered \cite{Buballa:2015awa}. Much less work has been devoted to the CDW in nuclear matter, where a few studies exist \cite{Dautry:1979bk, Heinz:2013hza}, however without systematic explorations of astrophysical relevance. Moreover, many of these studies were restricted either to simple approximations that completely ignored nucleonic vacuum fluctuations or plagued by artifacts induced by these fluctuations, which seemed to arise in conjunction with the CDW. Astrophysical applications and the theoretical issues were both addressed recently in a nucleon-meson model that has a well-defined chiral limit \cite{Pitsinigkos:2023xee,Papadopoulos:2024agt}. In these proceedings we review the main results of these two works. 

\section{Model: Free energy and parameter fit}

The nucleon-meson model we work with contains neutrons and protons as well as scalar and vector mesons and Yukawa interactions between the nucleons and  mesons. Details can be found for instance in Refs.\ \cite{Pitsinigkos:2023xee,Papadopoulos:2024agt,Drews:2014spa,Fraga:2018cvr}, with extensions to include strangeness discussed in Ref.\ \cite{Fraga:2022yls}. 
The CDW ansatz is implemented in the expectation values of the sigma and the neutral pion, 
\begin{equation}
\label{cdw}
\langle \sigma \rangle  = \phi \cos (2 \vec{q} \cdot \vec{x}) \, , \qquad  \langle \pi^0 \rangle = \phi \sin (2 \vec{q} \cdot \vec{x}) \, , 
\end{equation}
where the modulus $\phi$ and the wave number $q=|\vec{q}|$ need to be determined dynamically. Besides these condensates we also allow for nonzero expectation values of the temporal components of the omega and the rho meson, $\langle \omega_0 \rangle$ and $\langle \rho_0^3\rangle$. We work at zero temperature and neglect all mesonic fluctuations. The CDW induces anisotropic dispersion relations of the nucleons by breaking the spin degeneracy, 
\begin{equation}
    E_k^\pm = \sqrt{\left(\sqrt{k_\ell^2 + M^2} \pm q \right)^2 + k_\perp^2 } \, , 
\end{equation}
 where $k_\ell$ and $k_\perp$ are longitudinal and transverse components of the nucleon momentum with respect to the CDW wave vector $\vec{q}$, and where $M=g_\sigma \phi$ is the effective nucleon mass, with the Yukawa coupling constant $g_\sigma$. In this model, the entire nucleon mass is generated by the chiral condensate. While this is an over-simplified picture compared to QCD, it allows us to connect nuclear matter to a chirally restored phase at high densities. The vector meson condensates 
give rise to effective chemical potentials of the nucleons 
\bea
    \mu_n^* \equiv \mu_n - g_\omega \langle \omega_0 \rangle - g_\rho \langle \rho_0^3 \rangle \, , \qquad 
    \mu_p^* \equiv \mu_p - g_\omega \langle \omega_0 \rangle + g_\rho \langle \rho_0^3 \rangle \, ,
\eea
where $\mu_n$ and $\mu_p$ are neutron and proton chemical potentials. 
The free energy density can be written as 
\begin{equation}
\label{free}
    \Omega = \Omega_{\mathrm{nuc}} + \tilde{U} +\Delta \tilde{U} +  V + \Omega_{\mathrm{lep}} \, , 
\end{equation}
where we have added the free energy density of electrons and muons $\Omega_{\rm lep}$, needed in  the neutron star context. The nucleonic part 
(without Dirac sea) is 
\begin{equation}
    \Omega_{\mathrm{nuc}} = -\frac{1}{2\pi^2} \sum_{N = n,p} \sum_{s = \pm} \int_0^{\infty} d k_\ell  \int_0^{\infty} d k_\perp  k_\perp  (\mu_N^* - E_k^s) \Theta(\mu_{N}^* - E_k^s)   \, .
\end{equation}
The contribution from the Dirac sea can be combined with the scalar potential to 
\bea
    \tilde{U} & =& \sum_{n=1}^4 \frac{a_n}{n!}\frac{(\phi^2 - f_{\pi}^2)^n}{2^n} - \epsilon (\phi - f_{\pi}) \non[2ex]
    &&+ \frac{m_N^4}{96 \pi^2}\left(1 - 8 \frac{\phi^2}{f_{\pi}^2} - 12 \frac{\phi^4}{f_{\pi}^4} \ln \frac{\phi^2}{f_{\pi}^2} + 8 \frac{\phi^6}{f_{\pi}^6} - \frac{\phi^8}{f_{\pi}^8}\right),
\eea
with the pion decay constant $f_\pi$, parameters $a_1,\ldots,a_4$ and $\epsilon$ (the latter giving rise to a nonzero pion mass), and the vacuum mass of the nucleon $m_N$. All parameters are renormalized at this point. For the details of the renormalization procedure (with the help of proper time regularization), starting from the underlying Lagrangian with  bare parameters, see Ref.\ \cite{Pitsinigkos:2023xee}. We have separated the part that is only nonzero in the presence of a CDW, 
\begin{equation}
    \Delta \tilde{U} = 2 \phi^2 q^2 + (1 -\delta_{0q})\epsilon \phi - \frac{q^2 M^2}{2 \pi^2}\ln\frac{M^2}{\ell^2} - \frac{q^4}{2\pi^2}F(M/q),
\end{equation}
where we have abbreviated 
\begin{equation}
   F(y) \equiv \frac{1}{3} + \Theta(1 - y) \left[2 y^2 \left(1 + \frac{y^2}{4} \right)\ln \frac{1 + \sqrt{1 - y^2}}{y}- \sqrt{1 - y^2} \frac{2 + 13 y^2}{6}  \right] \, , 
\end{equation}
and where $\ell$ is a renormalization scale, which we choose to have the form \cite{Pitsinigkos:2023xee}
\begin{equation} \label{ell}
    \ell = \sqrt{m_N^2 + (2 c q)^2 } \, .
\end{equation}
For our results we will choose the numerical constant  
$c = 1/\exp(1 + 2\pi^2/g_\sigma^2) \simeq 0.3$, 
which is the limit below which the effective potential becomes unbounded with respect to $q$. Curing this unboundedness is the main reason for the choice (\ref{ell}), which also cures unphysical  phenomena of the CDW at large densities for $c=1$ \cite{Pitsinigkos:2023xee}. Choosing the smallest value  for $c$ renders the CDW as energetically favorable as possible, but re-introduces an unphysical boundary of the anisotropic phase at high densities. This choice makes our main result -- the absence of the CDW in neutron stars --  robust in the sense that no other choice of $c$ that respects boundedness of the potential in the $q$ direction will change our conclusion. Finally, the vector meson potential in Eq.\ (\ref{free}) is  
\begin{equation}
V = - \frac{m_\omega^2}{2}\langle \omega_0 \rangle^2 - \frac{m_\rho^2}{2}\langle \rho_0^3 \rangle^2 - \frac{d_\omega}{4}\langle \omega_0 \rangle^4 - \frac{d_\rho}{4}\langle \rho_0^3 \rangle^4 - \frac{d_{\omega \rho}}{2} \langle \omega_0 \rangle^2 \langle \rho_0^3 \rangle^2 \, , 
\end{equation}
with the vector meson masses $m_\omega$, $m_\rho$ and the coupling constants $d_{\omega}, d_{\rho}$, $d_{\omega \rho}$. 

Neutron star conditions are imposed by, firstly, requiring equilibrium with respect to electroweak processes. This is ensured by the condition $\mu_n = \mu_p + \mu_e$ and equality between the electron and muon chemical potentials, $\mu_e = \mu_\mu$. Secondly, we impose electric charge neutrality, which yields a condition on the number densities of protons, electrons, and muons, $n_p = n_e + n_\mu$. For a given neutron chemical potential $\mu_n$, we determine $\mu_e$ and the dynamical quantities $\phi$, $q$, $\la\omega_0\ra$, $\la\rho_0^3\ra$ from solving simultaneously the charge neutrality condition together with stationarity conditions of the free energy $\Omega$ with respect to the dynamical quantities. 

The model parameters are chosen to reproduce vacuum properties and  properties of isospin-symmetric nuclear matter at saturation, see Ref.\ \cite{Papadopoulos:2024agt} for a detailed explanation of the fitting procedure. Two of the quantities used for the fit, the (Dirac) mass of the nucleon at saturation $M_0$ and the incompressibility $K$, have significant empirical uncertainties. In the main results we shall choose $K=300\, {\rm MeV}$, which is at the upper boundary (perhaps somewhat beyond) of the empirical range, following a similar logic as explained for the choice of the constant $c$: as will become clear from the results, our main qualitative conclusions are robust against making nuclear matter softer by choosing a smaller value for $K$. The mass $M_0$ will be varied to explore the parameter space of the model. (When $M_0$ is varied, the parameters of the model are adjusted such that the physical properties of the vacuum and at saturation are maintained.) 

\begin{figure} [t]
\begin{center}
\includegraphics[width=0.5\textwidth]{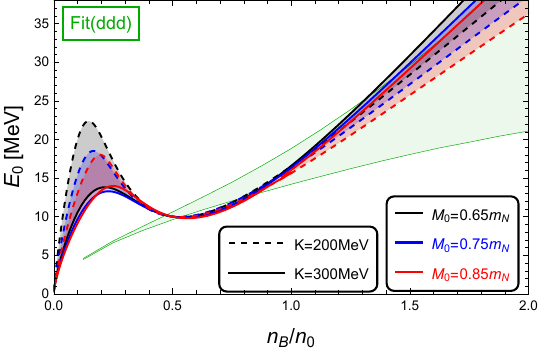}\includegraphics[width=0.5\textwidth]{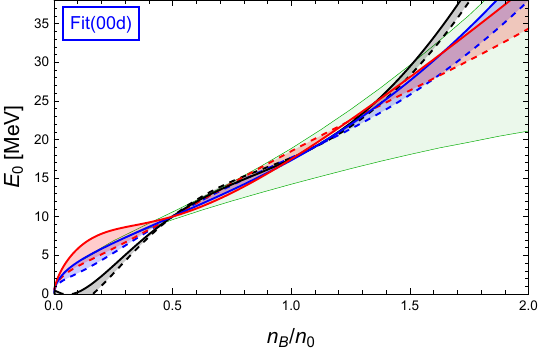}
\caption{Binding energy of pure neutron matter as a function of baryon density $n_B$ normalized to saturation density $n_0$, compared with 
results from chiral effective field theory given by the green error band \cite{Tews:2018kmu}. }   
\label{fig:bind}
\end{center}
\end{figure}

Importantly, we also employ pure neutron matter to fix our parameters
\cite{Alford:2022bpp}. This concerns mostly the choice of the vector meson self-couplings. We shall work with two different fits,
   \begin{subequations} \label{fits}
   \bea
        &&\mathrm{Fit(ddd):} \qquad  d \equiv d_\omega = d_\rho = d_{\omega \rho} \, .\\[2ex]
        &&\mathrm{Fit(00d):} \qquad d_\omega = d_\rho = 0, \qquad  d \equiv d_{\omega \rho} \, .
        \eea
    \end{subequations}
Their different effect on pure neutron matter is shown in Fig.\ \ref{fig:bind}, compared with results from chiral effective theory  
\cite{Tews:2018kmu}. Using the single degree of freedom $d$ from  each of the fits (\ref{fits}) to reproduce the binding energy $E_0|_{0.5 n_0} \simeq 10$ MeV, we see that only Fit(00d) is able to match the binding energy over a large density range, excellent agreement with effective field theory being reached for $M_0 = 0.75 m_N$.

\section{Results: CDW in neutron stars?}

\begin{figure}[t]
\begin{center}
\hbox{\includegraphics[width=0.5\textwidth]{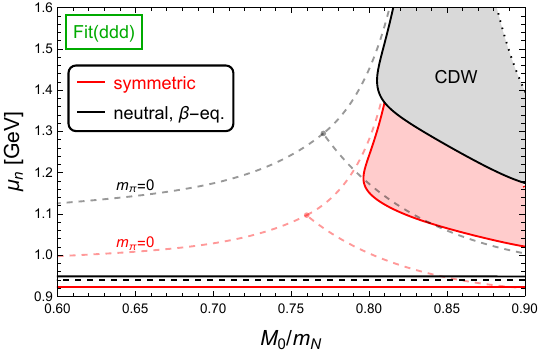}\includegraphics[width=0.5\textwidth]{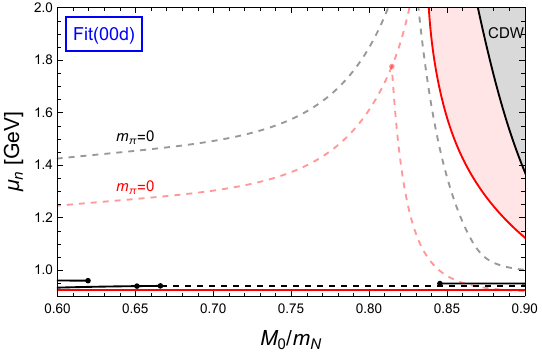}}
\caption{Zero-temperature phases in the plane of the neutron chemical potential $\mu_n$ and the nucleon mass at saturation $M_0$, which serves to scan the parameter space of the model. The figure compares the two fits (\ref{fits}) (left vs.\ right), and isospin-symmetric matter (red) with neutron star matter (black). Also, for comparison, the chiral limit (pale curves) is shown. Solid (dashed) lines are phase transitions of first (second) order.}   
\label{fig:phase}
\end{center}
\end{figure}

Let us first study the effect of the neutron star conditions on the fate of the CDW. This is done in Fig.\ \ref{fig:phase}. In the chiral limit of vanishing pion mass $m_\pi = 0$, we see that the CDW phase may not exist at all (towards small $M_0$) or appears in an intermediate region according to the schematic expectation of Fig.\ \ref{fig:QCDpd} (towards large $M_0$), with this intermediate region increasing as $M_0$ is increased. This trend is maintained in the case of a physical pion mass, where the phase transition without CDW is a crossover as a consequence of including the Dirac sea \cite{Pitsinigkos:2023xee}, and where the transition from isotropic nuclear matter to the CDW is of first order. There is a significant difference between the two fits (\ref{fits}), with Fit(ddd) being more favorable for the CDW. Most importantly, we see that neutron star conditions -- similar to the effect of the pion mass -- render the CDW less preferred by shifting its onset to larger chemical potentials.
In the left panel (upper right corner) we observe the unphysical boundary of the CDW mentioned above (where, in the present approach no other phase takes over consistently). A different renormalization scale, using for example $c=1$, would fix this, but make the region where the CDW is preferred even smaller. 

Less relevant for the main results, Fig.\ \ref{fig:phase} also shows different behaviors in the vicinity of the nuclear matter onset (bottom of both panels). While the onset of symmetric nuclear matter is of first order by construction (red solid horizontal lines), the onset for the case of $\beta$-equilibrated, charge neutral nuclear matter can be continuous or discontinuous. For Fit(ddd) a continuous onset is always followed by a first-order transition within nuclear matter, while for Fit(00d) different scenarios can appear as $M_0$ is varied. The empirically most realistic regime, $M_0\simeq (0.7-0.8)\, m_N$, shows a second-order onset without further phase transition afterwards.

\begin{figure}[t]
\includegraphics[width=\textwidth]{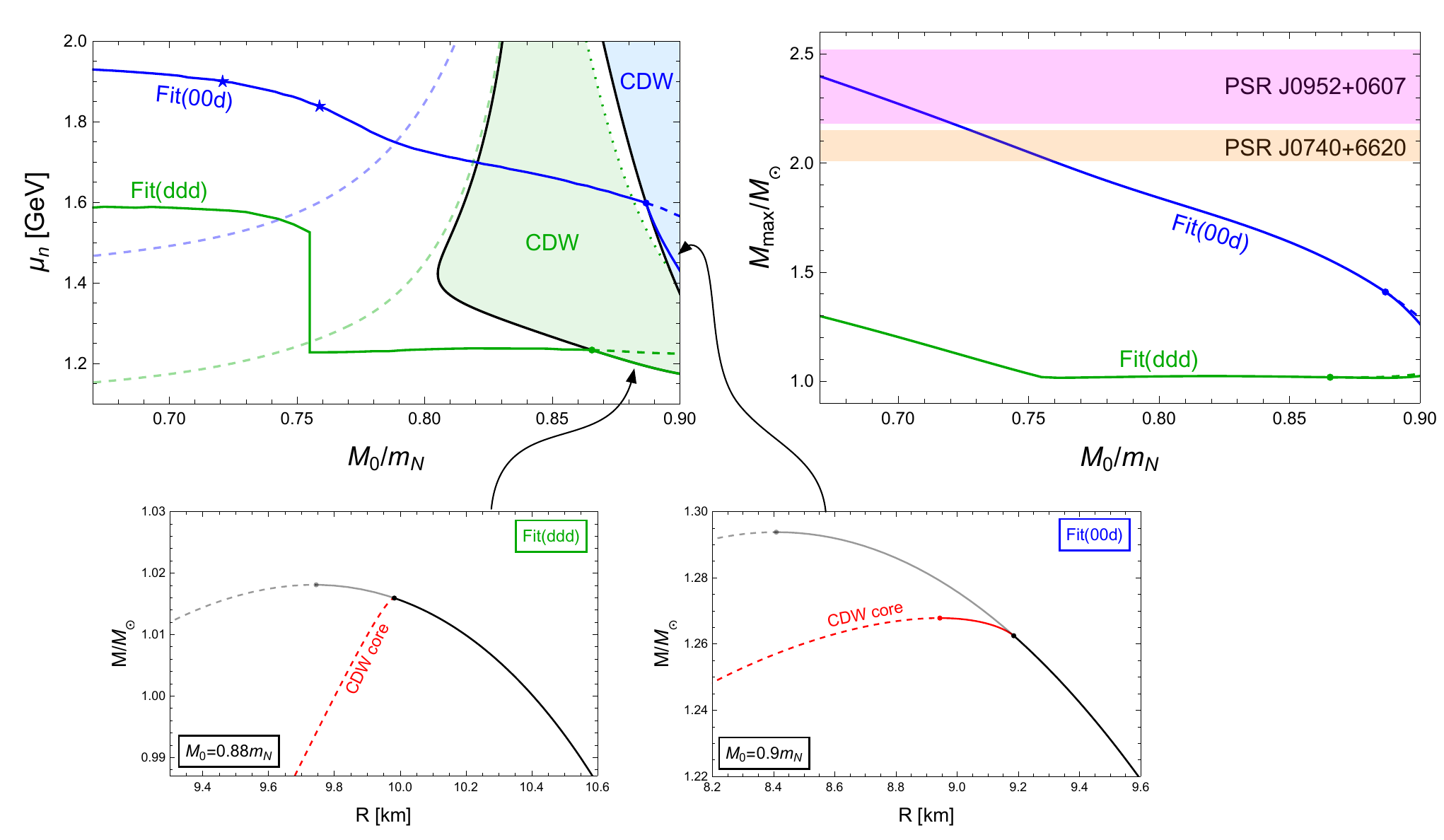}
\caption{
{\it Upper left panel:} Neutron chemical potential in the center of the maximally massive star for both fits, together with the CDW regions from Fig.\ \ref{fig:phase}. 
Stars with a CDW core exist only in an unstable branch of the mass-radius curve for Fit(ddd), while stable stars with a CDW core are possible for Fit(00d) (see lower panels). The gray segments in the lower panels as well as the dashed segments within the CDW regions in the upper panel are obtained, for comparison, by ignoring the CDW. 
{\it Upper right panel:} Maximum mass of the star in units of the solar mass $M_\odot$ for both fits compared to astrophysical data (orange and pink bands). Asterisks in the upper left panel indicate where the lower boundary of the uncertainty bands of these stars are reached. Only Fit(00d) can reproduce realistic stars. Stars with CDW cores belong to mass-radius curves that do not meet astrophysical constraints.
}   
\label{fig:stars}
\end{figure}

In order to construct neutron stars, we first compute the equation of state from pressure and energy density, straightforwardly obtained from our free energy density. The result is then used in the Tolman-Oppenheimer-Volkov equations, which we solve numerically to find the mass-radius curves. We assume the pressure to be isotropic, even in the presence of a CDW and we construct the entire star from our model, without adding a crust in the outer layers. This induces an error for the radii of our stars, but is sufficient for our purposes. For each  $M_0$ we compute the maximum mass of the star and the chemical potential  at the centre of the most massive star. The results are shown in Fig.\ \ref{fig:stars}. The upper left plot shows that two different scenarios are possible when the matter inside the star ``wants'' to develop a CDW: For Fit(ddd), the maximal $\mu_n$ coincides with the CDW onset curve because as soon as a CDW core develops, the star becomes unstable with respect to radial oscillations, as shown in the mass-radius curve in the lower left panel. The lower right panel shows that for Fit(00d) stable stars with CDW core exist. As a consequence, the maximal $\mu_n$ curve lies inside the (blue) CDW region. 

In the upper right panel, the mass of the most massive star for both  fits is compared with observations of PSR J0740+6620 \cite{Fonseca:2021wxt} and PSR J0952+0607 \cite{Romani:2022jhd}, two of the heaviest observed compact stars. The only stable stars with a CDW core are given by the segment on the (blue) Fit(00d) curve to the right of the dot. They are far from meeting the astrophysical constraints. For Fit(ddd), we see that for all $M_0$ considered here the equation of state is too soft to create sufficiently heavy stars (despite using an incompressibility at the upper end of the empirical range). In contrast, Fit(00d) does produce realistically heavy stars, which validates the model, albeit in a regime that does not exhibit a CDW. (For a specific set of model parameters needed to create two-solar-mass stars, for example with  $M_0=0.76\, m_N$, see Ref.\ \cite{Papadopoulos:2024agt}.) These realistic stars are hybrid in the sense that they contain a crossover to chirally restored matter. This is suggested by the pale dashed lines in the upper left panel, which   indicate the chiral transition in the chiral limit. The crossover has a particularly striking effect in the $\mu_n$ curve of Fit(ddd), where the discontinuity results from a twin star configuration, i.e., a mass-radius curve that shows stars with two different radii for the same mass (a feature usually assumed to be only possible in the presence of  first-order transitions).

\section{Summary and outlook}

We have investigated the CDW in nuclear matter and in the transition region to chirally restored matter. Our main results are, firstly, that neutron star conditions -- weak equilibrium and electric charge neutrality -- disfavor the CDW. Secondly, the CDW is preferred in a corner of the parameter space of our model that only allows for soft equations of state (soft even in the absence of the CDW) -- too soft to produce realistic neutron stars of at least two solar masses. We have shown that this is not a failure of the model itself, which does produce realistic neutron stars, in a parameter regime where the CDW does not exist. 

An obvious question for future work is whether a magnetic field can stabilize the CDW, as suggested by previous work in other models 
\cite{Ferrer:2019zfp,Gyory:2022hnv}, and if yes whether the necessary magnetic field strengths are realistic for neutron star interiors. Also, one can include the competing/coexisting effect of Cooper pairing and allow for phases that break translational invariance. One should also check the model dependence of our results, for instance by employing  the  extended linear sigma model of Refs.\ \cite{Heinz:2013hza,Haber:2014ula,Takeda:2018ldi}. Our results depend strongly on whether the nucleonic vacuum fluctuations are taken into account, and neither completely dropping them (previous work \cite{Heinz:2013hza}) as well as including them without further improvements of the approach (work reviewed here) are controlled approximations due to the large values of the Yukawa couplings. A useful complementary approach would therefore be to include the CDW in a controlled strong-coupling calculation based on the gauge-gravity duality, where it is known how to describe nuclear matter (with and without isospin-asymmetry) \cite{Kovensky:2021ddl,Ecker:2025sjb}.

\bibliographystyle{elsarticle-num}
\bibliography{references}

\end{document}